\documentstyle[12pt,aps]{revtex}
\tighten
\begin{document}
\title {Complete Set of Splitting Functions  Relevant in the Evolution of
Nucleonic Helicity Distributions}
\author{Rajen Kundu}
\address{ Saha Institute of Nuclear Physics \\
Sector I, Block AF, Bidhan Nagar,
Calcutta 700064 India}
\date{August 25, 1999}
\maketitle
\begin{abstract}
In this work, we show how the complete set of splitting
functions relevant for  the evolution of various distribution functions
describing nucleonic helicity structure can be obtained in the light front
Hamiltonian perturbation theory using completely fixed light front gauge, 
$A^+=0$.    
\end{abstract}
\section{Introduction}
How the total helicity of a nucleon is distributed among its constituents
is an active area of research at present. In a previous work\cite{rak}
\footnote{Henceforth referred, frequently in this work to avoid repetition,   
as paper I.}, 
we have shown that in a {\it gauge fixed} theory (with $A^+=0$ gauge) total
helicity operator can be satisfactorily defined and separated into orbital
and intrinsic parts for quarks and gluons and  defined the structure
functions containing information regarding the orbital helicity distribution
of the constituents. There we also employed our newly proposed methods of
calculating dressed parton structure functions in light front Hamiltonian
framework\cite{zhari} and obtained some of the splitting functions which are
relevant for the $Q^2$-evolution of these structure functions and the
corresponding anomalous dimensions, which were identical with the recent
results\cite{schaf} using different techniques (but with the same $A^+=0$
gauge). In this work, we shall use our gauge fixed formulation and
find out the complete set of splitting functions relevant in the evolution
of various helicity distributions to complete the picture. 

In our previous calculation of the various structure functions in paper I, 
we used the
dressed quark and gluon targets which are eigenstates of 
$P \equiv (P^+,P^1,P^2)$ and $S^3$. It was implicitly assumed there 
that the total
$P^\perp\equiv (P^1,P^2)$ for the targets to be zero, 
which one could always do 
by appropriately orienting the 3-axis. This in turn forced the orbital 
motion of the dressed parton associated with its centre of mass motion 
to vanish.
Consequently, our calculation produced the restricted number of splitting
functions which are important for the QCD evolution of the assumed state. To
obtain the complete set of relevant splitting functions and the
corresponding anomalous dimensions\cite{schaf} which,
for example, takes into account how the orbital motion evolves under QCD
interaction, one has to consider a target state where  orbital motion of the
constituents is
present even in absence of QCD interaction. This is nothing but due to 
the centre of mass motion of the dressed parton.  Note that this centre of
mass motion has nothing to do with that of a composite system such as meson. 
For the calculation involving a meson target, it 
will be replaced by the internal motion of the concerned parton. 
We shall introduce the relevant target state for our calculation
in  
Sec.II, where we have also introduced the multi-parton wave-functions
relevant for our calculations.  In Secs.III-IV we present our calculations
and then conclude in Sec.V. 
\section{The target state and multi-parton wave-functions}
Required state for the dressed parton can be obtained by taking a suitable
superposition of states where all the individual states carry definite
longitudinal and transverse momenta ($P$) as well as the intrinsic helicity
($\sigma$) and then projecting out those states which carry same $j^3$ by
introducing a Kronecker-delta function. We consider the following as our
target state
\begin{equation}
\mid S\rangle 
= \sum_{\sigma} \sum_m e^{im\phi} \Phi_{\rm cm}
^{j^3}(P^+,\mid
P^{\perp}\mid, m)\delta_{m, j^3-\sigma}\mid P,\sigma\rangle\, , 
\label{lcm}
\end{equation}
where $\phi$ parametrizes the state such that
$(P^1,P^2)=|P^{\perp}|(\cos\phi, \sin\phi)$. Note that we have written the
$\phi$-dependence separately as a sum over $m$, the eigenvalue of $L^3$, so
that  $\Phi_{\rm cm}$ is independent of $\phi$.  
The dressed parton state
$|P,\sigma\rangle$ in the RHS of eq.(\ref{lcm}) can be written in the
Fock-basis using multi-parton wave-functions which are independent of
momenta $P$ \cite{zhang93}. 
For a dressed quark, it is given by 
\begin{eqnarray}
	|P^+,P_\bot,\sigma \rangle &=& 
		\sqrt{{\cal N}_q}\Big\{ b^\dagger
		(k,\sigma)|0\rangle + \sum_{\sigma_1\lambda} \int {dk_1^+d^2
		k_{\bot 1}\over \sqrt{2(2\pi)^3 k_1^+}} 
                {dk_2^+d^2k_{\bot 2}\over \sqrt{2 (2\pi)^3k_2^+}} 
     \sqrt{2(2\pi)^3 P^+} \delta^3(P-k_1-k_2) \nonumber \\
	& &~~~~~~~~~~~~~~~~~~~~~~~~~~~~~~~\times \Phi^{\sigma}_{\sigma_1
		\lambda}(P;k_1,k_2)b^\dagger
		(k_1,\sigma_1) a^\dagger (k_2,\lambda) | 0 \rangle + 
		\cdots \Big\},\label{dsqs}\, ,
\end{eqnarray}
where the normalization constant ${\cal N}_q$ is  determined perturbatively
from the normalization condition, 
\begin{equation}
	\langle {P'}^+,P'_\bot,\sigma' |P^+,P_\bot,\sigma \rangle
	= 2(2\pi)^3 P^+ \delta_{\sigma,\sigma'}\delta(P^+-{P'}^+)
	\delta^2(P_\bot-P'_\bot). \label{norm}\, .
\end{equation}
Similarly for a dressed gluon
\begin{eqnarray}
	|P^+,P_\bot,\lambda \rangle &=& 
		\sqrt{{\cal N}_g}\Big\{ a^\dagger
		(k,\sigma)|0\rangle + \sum_{\sigma_1\sigma_2} \int {dk_1^+d^2
		k_{\bot 1}\over \sqrt{2(2\pi)^3 k_1^+}} 
                {dk_2^+d^2k_{\bot 2}\over \sqrt{2 (2\pi)^3k_2^+}} 
     \sqrt{2(2\pi)^3 P^+} \delta^3(P-k_1-k_2) \nonumber \\
	& &~~~~~~~~~~~~~~~~~~~~~~~~~~~~~~~\times \Phi
		^{\lambda}_{\sigma_1\sigma_2}(P;k_1,k_2)b^\dagger
		(k_1,\sigma_1) d^\dagger (k_2,\sigma_2) | 0
	\rangle\nonumber\\ & &~~~~~~~~~~~~~~~~ + 
		{1\over 2}\sum_{\lambda_1\lambda_2} \int {dk_1^+d^2
		k_{\bot 1}\over \sqrt{2(2\pi)^3 k_1^+}} 
                {dk_2^+d^2k_{\bot 2}\over \sqrt{2 (2\pi)^3k_2^+}} 
     \sqrt{2(2\pi)^3 P^+} \delta^3(P-k_1-k_2) \nonumber \\
	& &~~~~~~~~~~~~~~~~~~~~~~~~~~~~~~~\times \Phi
		^{\lambda}_{\lambda_1\lambda_2}(P;k_1,k_2)a^\dagger
		(k_1,\lambda_1) a^\dagger (k_2,\lambda_2) | 0 \rangle + 
		\cdots \Big\}\, \label{dsgs} .
\end{eqnarray}

We introduce the boost invariant amplitudes  
$ \psi^\sigma_{\sigma_1,
\lambda_2}(x,\kappa^\perp)$ and so on by  
$\Phi^\sigma_{\lambda_1\lambda_2}(P;k_1,k_2)  
= { 1 \over \sqrt{P^+}}  \psi^\lambda_{\sigma_1
\lambda_2}(x,\kappa^\perp)$ and so on. Here the relative momenta
($x,\kappa^\perp$) are defined
as
\begin{equation}
k^+_1=xP^+,~k^i_1=\kappa^i + x P^i ~~ {\rm and} ~~
k^+_2=(1-x)P^+,~k^i_2=-\kappa^i +(1-x) x P^i \, .
\end{equation}
From the light-front QCD Hamiltonian, to the lowest
order in perturbation theory, we have \cite{zhari},
\begin{eqnarray}
&&	\psi^{\sigma\,~|qg\rangle}_{\sigma_1\lambda_2}(x,\kappa_\bot) = - 
	{g \over \sqrt{2 (2 \pi)^3}}
		T^a {x(1-x)\over \kappa_\bot^2 }
		\chi^\dagger_{\sigma_1} \Bigg\{2{\kappa_\bot^i \over
		1-x} 
+ {1\over x}({\sigma_\bot}\cdot \kappa_\bot)
		{\sigma}^i \Bigg\}
		\chi_\sigma \varepsilon^{i*}_{\lambda_2}\, ,
 \label{psip}\\
&&	\psi^{\lambda~|q\bar{q}\rangle}_{\sigma_1\sigma_2}
	(x,\kappa_\bot) =~ - 
	{g \over \sqrt{2 (2 \pi)^3}}
		T^a {x(1-x)\over \kappa_\bot^2 }
		\chi^\dagger_{\sigma_1} \Bigg\{{({\sigma_\bot}\cdot
		\kappa_\bot) \over x}\sigma_\bot^i 
		 - \sigma_\bot^i{({\sigma_\bot}\cdot \kappa_\bot)
		\over 1-x} \Bigg\}
		\chi_{-\sigma_2} \varepsilon^{i}_{\lambda}\, ,
\\
&&	\psi^{\lambda~|gg\rangle}_{\lambda_1\lambda_2}(x,\kappa_\bot) = - 
	{2igf^{abc} \over \sqrt{2 (2 \pi)^3}}~{\sqrt{x(1-x)}\over \kappa^2}
	~\Bigg\{-\kappa^i~\delta_{jl}~+~{\kappa^j\over x}\delta_{il}
	~+~{\kappa^l\over 1-x}\delta_{ij}~\Bigg\}~ 
	\varepsilon^{j*}_{\lambda_1}\varepsilon^{l*}_{\lambda_2}
	\varepsilon^{i}_{\lambda}\, .
\end{eqnarray}
Here we have already taken target mass and bare quark mass to be zero, 
$M=m=0$, since nonzero masses in the above
wave-functions are going to produce higher twist effects. Our next task is
to calculate the structure functions with $\mid S \rangle$ being the dressed
quark target [{\it i.e.}, eq.(\ref{lcm}) 
and eq.(\ref{dsqs}) combined] and dressed
gluon target [{\it i.e.}, eq.(\ref{lcm}) and eq.(\ref{dsgs}) combined]. 

\section{Dressed Quark Structure Functions}
For the dressed quark eq.(\ref{lcm}) can be written as
\begin{equation}
|S\rangle ~=~ e^{im_1\phi}~\Phi_{\rm cm} (m_1)~|P,+{1\over 2}\rangle ~+~
	    e^{im_2\phi}~ \Phi_{\rm cm}(m_2)~|P,-{1\over 2}
	    \rangle\, ,\label{qqlcm}
\end{equation}
where $m_1=j^3-{1\over 2}$ and $m_2=j^3+{1\over 2}$. 
For such a state,  in the zeroth
order perturbation theory, {\it i.e.}, if the two-particle wave-function in
eq.(\ref{dsqs}) is zero,  $\Delta q(x,Q^2)$ and $\Delta
q_L(x,Q^2)$  
come out to be proportional to $\delta$-functions and are  
given by\footnote{ See paper I for the definitions of various structure
functions.}
\begin{eqnarray}
&&\Delta q(x,Q^2)= {{\cal N}_q\over 2}\left\{~
                        |\Phi_{\rm cm} (m_1)|^2
                        -|\Phi_{\rm cm}(m_2)|^2~
\right\}\delta(1-x)\equiv {1\over 2} \Delta \Sigma~\delta(1-x)\, , \nonumber\\
&&\Delta q_L(x,Q^2)={\cal N}_q\left\{~
                	m_1~|\Phi_{\rm cm} (m_1)|^2
       			+m_2~|\Phi_{\rm cm}
			(m_2)|^2 ~\right\}\delta(1-x)\equiv L_q~\delta(1-x)\, ,
\end{eqnarray}
while $\Delta g(x,Q^2)$ and $\Delta g_L(x,Q^2)$ are zero. Of course, ${\cal
N}_q=1$ has been used above, 
since calculations are performed only in zeroth order 
in the coupling.

Now, the structure functions can be calculated to the first order in
light-front perturbation theory by truncating the dressed quark state in
eq.(\ref{dsqs}) at
the two particle level and using the wave-function in eq.(\ref{psip}). 
Straight-forward calculations give 
\begin{equation}
\Delta q(x,Q^2) = {{\cal N}_q\over 2}~\Delta \Sigma~ 
			\Big\{~ \delta(1-x) + {\alpha_s\over 2\pi}C_f \ln
			{\Lambda^2\over \mu^2}~ {1+x^2\over 1-x}~\Big\}\, ,
\label{qsl}
\end{equation}
where $\Lambda$ and $\mu$ are the upper and lower momentum cutoff.
Note that it is same as we obtained in paper I 
except an extra weight factor
$\Delta \Sigma$ 
coming from the fact that the target state now is a particular superposition. 
Similarly,
\begin{equation}
\Delta g(1-x,Q^2) = {\cal N}_q~\Delta \Sigma ~
			{\alpha_s\over 2\pi}C_f \ln
			{\Lambda^2\over \mu^2}~(1+x)\, .
\label{gsl}
\end{equation}
On the other hand, we get extra terms while calculating $\Delta q_L(x,Q^2)$
and $\Delta g_L(1-x,Q^2)$ as given below.
\begin{eqnarray}
&&\Delta q_L(x,Q^2) = -~{\cal N}_q ~\Delta \Sigma~ 
			{\alpha_s\over 2\pi}C_f \ln
			{\Lambda^2\over \mu^2}~ (1-x)(1+x)\nonumber\\
&&~~~~~~~~~~~~~~~~~~~~~~~~~~~~~~+ ~{\cal N}_q~L_q~\Big\{~\delta(1-x)+
			{\alpha_s\over 2\pi}C_f \ln    	  
                        {\Lambda^2\over \mu^2}~ {x^3+x\over 1-x}~\Big\}\, 
\label{qll}
\\
&&\Delta g_L(1-x,Q^2) = -~{\cal N}_q~\Delta \Sigma~
			{\alpha_s\over 2\pi}C_f \ln
			{\Lambda^2\over \mu^2}~ x(1+x)\nonumber\\
&&~~~~~~~~~~~~~~~~~~~~~~~~~~~~~~+~{\cal N}_q~ L_q~
			{\alpha_s\over 2\pi}C_f \ln         
			{\Lambda^2\over \mu^2}~ (1+x^2) \, 
\label{gll}
\end{eqnarray}
The extra terms in eq.(\ref{qll}) and eq.(\ref{gll}) are coming due to the
fact that we have used\cite{rak} 
\begin{eqnarray}
&& \hat{L}^3_q=-i\Big[ ~(1-x){\partial 
\over \partial \theta} +x{\partial \over 
\partial \phi}~\Big]\nonumber\\
&& \hat{L}^3_g=-i\Big[~ x{\partial \over 
\partial \theta} +(1-x){\partial \over 
\partial \phi}~\Big]
\end{eqnarray}
where $\theta$ is defined by $(\kappa^1,\kappa^2)\equiv 
|\kappa^\perp|(\cos\theta,
\sin\theta)$. In our previous calculation ${\partial \over \partial \phi}$
did not contribute for we assumed $P^\perp=0$. Also notice that the other
terms in $\hat{L}^3_{q,g}$ (as given in paper I) have no contribution for
they produced odd integral of $\kappa^\perp$ over symmetric limit. 

Now, the normalization constant ${\cal N}_q$ to the order $\alpha_s$ can be
calculated from eq.(\ref{norm}) and is given by \cite{zhari}
\begin{eqnarray}
&&{\cal N}_q =~ \Big\{ 1 - {\alpha_s \over 2\pi} C_f 
		\ln {\Lambda^2 \over \mu^2}\int
		dx ~{1+x^2\over 1-x}\Big\}\, ,\nonumber\\
&&~~~~=~1~-~C_f~{\alpha_s\over 2\pi} \ln {\Lambda^2\over \mu^2}~
\Big[~\int dx ~{2\over 1-x} ~-~{3\over
2}~\Big]\, .
\end{eqnarray}
Putting ${\cal N}_q$ back into eqs.(\ref{qsl}-\ref{gll}) and keeping terms
only to the order $\alpha_s$, we have 
\begin{eqnarray}
&&\Delta q(x,Q^2) =~~ {1\over 2}\Delta \Sigma
			~\Big\{~\delta(1-x) + {\alpha_s\over 2\pi} \ln
			{\Lambda^2\over \mu^2} ~P_{SS(qq)}(x)~\Big\}\, ,
\nonumber\\
&&\Delta g(1-x,Q^2) =~~   \Delta \Sigma
			 ~{\alpha_s\over 2\pi} \ln
			{\Lambda^2\over \mu^2} ~P_{SS(gq)}(1-x)\, ,
\nonumber\\
&&\Delta q_L(x,Q^2)=~   \Delta \Sigma
			 ~{\alpha_s\over 2\pi} \ln
			{\Lambda^2\over \mu^2}~ P_{LS(gq)}(x)
\nonumber\\
&&~~~~~~~~~~~~~~~~~~~~~~~~~~~~~~~~+L_q ~ 	 
			\Big\{~\delta(1-x) + {\alpha_s\over 2\pi} \ln
			{\Lambda^2\over \mu^2} ~P_{LL(qq)}(x)
			~\Big\}\, ,\nonumber\\
&&\Delta g_L(1-x,Q^2)=~   \Delta \Sigma
			 ~{\alpha_s\over 2\pi} \ln
 			{\Lambda^2\over \mu^2} ~P_{LS(gq)}(1-x)
\nonumber\\
&&~~~~~~~~~~~~~~~~~~~~~~~~~~~~~~~~+~L_q~
			 {\alpha_s\over 2\pi} \ln
			{\Lambda^2\over \mu^2} ~P_{LL(gq)}(1-x)\, ,
\end{eqnarray}
where we have defined the various splitting functions as follows. (Notice
that the argument of glunic distribution functions is $(1-x)$ which is the
momentum fraction carried by the gluon here.) 
\begin{eqnarray}
&& P_{SS(qq)}(x)=~C_f~\Big[{1+x^2\over (1-x)_+} ~+~{3\over
2}\delta(1-x)\Big]\, ,
\nonumber\\
&& P_{SS(gq)}(1-x)=~C_f~(1+x) \nonumber\\
&& P_{LS(qq)}(x)  =~-C_f~(1-x^2)\nonumber\\
&& P_{LS(gq)}(1-x)=~-C_f~x(1+x)\nonumber\\
&& P_{LL(qq)}(x)=~C_f~\Big[{x^3+x\over (1-x)_+}~+~{3\over 2}\delta(1-x)
\Big]\, ,\nonumber\\
&& P_{LL(gq)}(1-x)=~C_f~(1+x^2)\label{jpl}
\end{eqnarray}  
Also, since there are no contributions coming from $L_q$ in $\Delta
q(x,Q^2)$ and $\Delta g(x,Q^2)$, we have,
\begin{equation}
P_{SL(qq)}(x)~=~P_{SL(gq)}(x)~=~0\, .\label{eee}
\end{equation}
Last two splitting functions in eq.(\ref{jpl}) and that in eq.(\ref{eee}) 
are new compared to those in paper I.

\section{Dressed Gluon Structure Functions}
Now we repeat the calculations given in the previous section by replacing
the target state with a dressed gluon. 
For the dressed gluon eq.(\ref{lcm}) can be written as
\begin{equation}
|S\rangle ~=~ e^{im_1\phi}~\Phi_{\rm cm} (m_1)~|P,+1\rangle ~+~
	    e^{im_2\phi}~ \Phi_{\rm cm}(m_2)~|P,-1
	    \rangle\, ,
\end{equation}
where $m_1=j^3-1$ and $m_2=j^3+1$. 
For such a state, in contrast to the quark case, $\Delta g(x,Q^2)$ 
and $\Delta g_L(x,Q^2)$ are proportional to the $\delta$-function 
in the zeroth
order perturbation theory while 
$\Delta q(x,Q^2)$ and $\Delta q_L(x,Q^2)$ are zero. 
\begin{eqnarray}
&&\Delta g(x,Q^2)= {\cal N}_g\left\{~
                        |\Phi_{\rm cm} (m_1)|^2
                        -|\Phi_{\rm cm}(m_2)|^2~
\right\}\delta(1-x)\equiv \Delta g~\delta(1-x)\, , \nonumber\\
&&\Delta g_L(x,Q^2)={\cal N}_g\left\{~
                	m_1~|\Phi_{\rm cm} (m_1)|^2
       			+m_2~|\Phi_{\rm cm}
			(m_2)|^2 ~\right\}\delta(1-x)\equiv L_g~\delta(1-x)\, ,
\end{eqnarray}

Now, with the dressed gluon state truncated at the two-particle level as
given in eq.(\ref{dsgs}), we can calculate all the structure functions to
first order in $\alpha_s$ and the results are as follows 
\begin{eqnarray}
&&\Delta q(x,Q^2)~~ = ~\Delta g~{\alpha_s\over 2\pi}    
ln{\Lambda^2\over \mu^2} ~P_{SS(qg)}(x)\nonumber\\
&&\Delta q_L(x,Q^2)~=~ \Delta g~{\alpha_s\over 2\pi} 
ln{\Lambda^2\over \mu^2} ~2P_{LS(qg)}(x)~+~L_g~{\alpha_s\over 2\pi} 
ln{\Lambda^2\over \mu^2} ~2P_{LL(qg)}(x) \, ,\nonumber\\
&&\Delta g(x,Q^2)~~= ~\Delta g~\Big\{~\delta(1-x)~+~{\alpha_s\over 2\pi} 
ln{\Lambda^2\over \mu^2} ~P_{SS(gg)}(x)~\Big\} \, , \nonumber\\
&&\Delta g_L(x,Q^2)~=~\Delta g~{\alpha_s\over 2\pi} 
ln{\Lambda^2\over \mu^2} ~P_{LS(gg)}(x)\nonumber\\
&&~~~~~~~~~~~~~~~~~~~~~~~~+~L_g~\Big\{~\delta(1-x) ~+~{\alpha_s\over 2\pi} 
ln{\Lambda^2\over \mu^2} ~P_{LL(gg)}(x) ~\Big\}\, 
\end{eqnarray}
where, like  the quark case, 
we have defined the  splitting functions as follows.
\begin{eqnarray}
&&P_{SS(qg)}(x)~=~{1\over 2}~[~x^2- (1-x)^2~]\, ,\nonumber\\
&&P_{SS(gg)}(x)~=~N~\Big\{(1+x^4)\big[{1\over x} + 
{1\over (1-x)_+}\big] 
-{(1-x)^3\over x}\Big\}~+~\delta(1-x)~\Big({11\over 6}N
~-~{1\over 3}\Big)\, ,\nonumber\\
&&P_{LS(qg)}(x)~=~{1\over 2}~[~x^2+ (1-x)^2~](1-x)\, ,\nonumber\\
&&P_{LS(gg)}(x)~=~2N~(x-1)(x^2-x+2)\, ,\nonumber\\
&&P_{LL(qg)}(x)~=~{1\over 2}~x~[~x^2+ (1-x)^2~]\, ,\nonumber\\
&&P_{LL(gg)}(x)~=~2N~x\Big\{{x\over (1-x)_+} 
+ {1-x\over x} +x(1-x)\Big\}
+\delta(1-x)~\Big({11\over 6}N~-~ {1\over 3}\Big)\, , 
\label{spg}
\end{eqnarray}
and 
\begin{equation}
P_{SL(qg)}(x)~=~P_{SL(gg)}(x)~=~0\, .\label{spg2}
\end{equation}
Here, in the above calculation, we have used the normalization constant
${\cal N}_g$ as \cite{zhari}
\begin{eqnarray}
&&{\cal N}_g~=~1- {\alpha_s\over 2\pi} \ln {\Lambda^2\over \mu^2} ~
\int dx ~\Big[~2Nx\big\{{x\over 1-x} + {1-x\over x} +
x(1-x)\big\}~+~{1\over 2}~[~x^2+ (1-x)^2~]\Big]\, ,\nonumber\\
&&~~~~~=~1~-~{\alpha_s\over 2\pi} \ln {\Lambda^2\over \mu^2}
\int dy~{2N\over 1-y}~+~\Big({11\over 6}N-
{1\over 3}\Big)\, .   
\end{eqnarray}
Last two splitting functions in eq.(\ref{spg}) and that in eq.(\ref{spg2})
are new compared to those in paper I. Also, to the best of our knowledge,
spin-dependent Altarelli-Parisi splitting functions denoted here as 
$P_{SS(qg)}$ and $P_{SS(gg)}$ are explicitly derived using light-front 
Hamiltonian perturbation theory for the first time here.

Corresponding anomalous dimensions are defined as 
\begin{equation}
A^n_{AB(ab)}~=~ \int dx x^{n-1}~P_{AB(ab)}(x)\, .
\end{equation}
Using this definition the results given in Ref.\cite{schaf} for the
anomalous dimensions are exactly reproduced and it is 
needless to rewrite them here. 
Note that for these dressed parton states, expectation value of $J^3$ to this
order in perturbation theory is independent of $\alpha_s$, as it should be
due to kinematical nature of the helicity operator,
\begin{eqnarray}
&&{\langle S|J^3| S\rangle_q \over 2(2\pi)^3 P^+\delta^3(0)} =~
			{1\over 2}~ \Delta\Sigma~
			+~L_q \, ,\nonumber\\
&&{\langle S| J^3|S\rangle_g\over 2(2\pi)^3 P^+\delta^3(0)}~
=~\Delta g ~+~ L_g\, .	 
\label{srul}
\end{eqnarray}			
Eqs.(\ref{srul}) explicitly verify the helicity sum rule for the two
different states considered here and show 
the consistency of our calculation. 
\section{Conclusion}
In this work, we have presented the calculations of the complete set of
splitting functions relevant in the evolution of the various helicity
distribution functions defined in the gauge $A^+=0$. This work in
association with our previous work in paper I gives a satisfactory
description of various parts of helicity distributions and their evolutions in
the {\it gauge fixed light front Hamiltonian formulation} and in conformity
with the work presented in Ref.\cite{schaf}. 
 
It should be noted that our formulation and results are obtained in a gauge
fixed theory (same is true for the calculation in Ref.\cite{schaf}).  
At present, we have very little idea in which processes the orbital helicity
distributions that we have defined  can be measured experimentally. 
It needs to be admitted that in absence of the
experimental support and/or our
results are reproduced by some other means, 
{\it the gauge fixed} formulation is  little unsettled. 
On the other hand, 
gauge invariance in separating the helicity operator into 
intrinsic and orbital parts for quarks and gluons is a long standing issue.
Lot of work has been done in the last few years towards defining such
operators showing full respect to the gauge invariance\cite{ji}. Also, 
there are
attempts to define operators compatible to the residual gauge invariance in
light-front gauge\cite{bj}. In spite of all these works,
the issue still seems to be a little unsettled\cite{dz} as well.  
In view of this dilemma, it needs thorough investigation to settle down the
question of gauge invariance and/or find out processes sensitive to OAM that
can be measured in the experiment. We hope to undertake such 
investigations in future.
\section*{Acknowledgement} 
I would like to thank Prof. A. Harindranath for a careful reading of the
manuscript and providing important suggestions. I would also like to
acknowledge 
useful discussions with my colleagues in Theory Division, SINP. 
\appendix
\section{}
In this appendix, we provide the necessary 
details of calculation in arriving at the
expression for $\Delta q_L(x,Q^2)$ given eq.(\ref{qll}) starting from its
definition as an example. 
All the other splitting functions can be obtained in a 
similar manner. We have (see paper I),
\begin{eqnarray}
\Delta q_L(x, Q^2) = { 1 \over 4 \pi P^+} \int d \eta e^{ -i  \eta x} 
\langle S \mid \Big [ {\overline \psi}(\xi^-) \gamma^+ i (x^1 \partial^2
- x^2 \partial^1) \psi(0) + h.c.\Big ] \mid S \rangle \label {fo}
\end{eqnarray}
with $ \eta = {1 \over 2} P^+ \xi^-$. 
Note that the operator involved here is a number conserving operator, so
that the  non-vanishing contributions will only be diagonal in particle number.
Let us first consider the contribution coming from the first term in
$|S\rangle$ given in eq.(\ref{qqlcm}). 
Using the Fourier expansion for the  dynamical
fermionic field $\psi^+$ and the standard commutation relation between
creation and annihilation operators (see paper I), we get two
non-vanishing 
contributions. First one coming from the one particle sector and is simply
given by
\begin{eqnarray}
\Delta q_L(1)\label{q1}
&&=~{\cal N}_q~
|\Phi_{\rm cm}(m_1)|^2 ~e^{-im_1\phi}(-i{\partial\over\partial \phi})e^{i
m_1\phi}~\delta(1-x)\nonumber\\
&&=~{\cal N}_q~m_1|\Phi_{\rm cm}(m_1)|^2~\delta(1-x)\,.
\end{eqnarray}
The second one coming from the two particle sector is given by
\begin{eqnarray}
\Delta q_L(2)\label{q2}
&&=~{\cal N}_q|\Phi_{\rm cm}(m_1)|^2~
\sum_{\sigma_1, \lambda_2}   ~\int d^2 \kappa^\perp
~e^{-im_1\phi}~
{\psi^{\uparrow}_{\sigma_1 \lambda_2}}^*(x, \kappa^\perp)\nonumber\\
&&~~~~~~~~~~~~~~~~~~ 
(-i)\Big[(1-x) {\partial \over \partial \theta}+ x{\partial \over \partial
\phi}\Big]
{\psi^{\uparrow}_{\sigma_1 \lambda_2}}(x, \kappa^\perp)~
e^{im_1\phi} \nonumber \\
&& =~{\cal N}_q|\Phi_{\rm cm}(m_1)|^2~\sum_{\sigma_1, \lambda_2}  
 ~\int d^2 \kappa^\perp
~(1-x){\psi^{\uparrow}_{\sigma_1 \lambda_2}}^*(x, \kappa^\perp) 
\left(-i {\partial \over \partial \theta}\right)
\psi^{\uparrow}_{\sigma_1 \lambda_2}(x, \kappa^\perp)~ \nonumber \\
&&~~~~~~~~~~~~~~~~+{\cal N}_q~m_1|\Phi_{\rm cm}(m_1)|^2~
\sum_{\sigma_1, \lambda_2}   ~\int d^2 \kappa^\perp
~x~{\psi^{\uparrow}_{\sigma_1 \lambda_2}}^*(x, \kappa^\perp) 
{\psi^{\uparrow}_{\sigma_1 \lambda_2}}(x, \kappa^\perp)\label{phm1}
\end{eqnarray}
Here and in the following $\uparrow$ implies helicity $+{1\over 2}$ for
quarks and $+1$ for gluons and so on. Using
$ \chi_{1\over 2}=(1, 0)$, $ \chi_{-{1\over 2}}=(0, 1)$  and 
$ \varepsilon_{\pm1}={1\over\sqrt 2}(1,\pm i)$ in eq.(\ref{psip}), various
components of the two-particle wave-function can be simplified and are given
by 
\begin{eqnarray}
&&~\psi^\uparrow_{\uparrow\uparrow}(x,\kappa^\perp)~=
~-{g T^a\over\sqrt{2(2\pi)^3}}{\sqrt{2}
e^{-i\theta}\over |\kappa^\perp|\sqrt{1-x}},~~~~~
\psi^\uparrow_{\uparrow\downarrow}(x,\kappa^\perp)~=
~-{g T^a\over\sqrt{2(2\pi)^3}}~{\sqrt{2}xe^{i\theta}\over 
|\kappa^\perp| \sqrt{1-x}}, \nonumber\\
&&~\psi^\uparrow_{\downarrow\uparrow}(x,\kappa^\perp)=
\psi^\uparrow_{\downarrow\downarrow}(x,\kappa^\perp)=0.
\end{eqnarray}
Last two components are zero since we have assumed $m_q=0$, which forces the  
quark helicity flip interaction to vanish. 
Putting them back into eq.(\ref{phm1}) and performing
$\kappa^\perp$-integration, we get the required result from the two-particle
sector. Thus the total contribution from the first term of $|S\rangle$
becomes
\begin{eqnarray}
\label{st1}
\Delta q^I_L(1) +\Delta q^I_L(2)~=~&&-~{\cal N}_q ~|\Phi_{\rm cm}(m_1)|^2~ 
			{\alpha_s\over 2\pi}C_f \ln
		{\Lambda^2\over \mu^2}~ (1-x)(1+x)\nonumber\\
&&~~~~~~~+ ~{\cal N}_q~m_1
   			|\Phi_{\rm cm}(m_1)|^2~\Big\{~\delta(1-x)+
			{\alpha_s\over 2\pi}C_f \ln    	  
                        {\Lambda^2\over \mu^2}~ {x^3+x\over 1-x}~\Big\}\, . 
\end{eqnarray}
The contribution from the second term in $|S\rangle$ is again given by
eq.(\ref{q1}) and eq.(\ref{q2}) with $m_1\rightarrow m_2$ and
$\uparrow\rightarrow\downarrow$. Then one uses
\begin{eqnarray}
&&~\psi^\downarrow_{\uparrow\uparrow}(x,\kappa^\perp)~=~
\psi^\downarrow_{\uparrow\downarrow}(x,\kappa^{\perp})~=~0,\nonumber\\ 
&&\psi^\downarrow_{\downarrow\uparrow}(x,\kappa^\perp)~=
~-{g T^a\over\sqrt{2(2\pi)^3}}{\sqrt{2}
xe^{-i\theta}\over|\kappa^\perp|\sqrt{1-x}},~~~~
\psi^\downarrow_{\downarrow\downarrow}(x,\kappa^\perp)~=
~-{g T^a\over\sqrt{2(2\pi)^3}}{\sqrt{2}
e^{i\theta}\over |\kappa^\perp|\sqrt{1-x}}, 
\end{eqnarray}
to obtain 
\begin{eqnarray}\label{st2}
\Delta q^{II}_L(1) +\Delta q^{II}_L(2)~=~
&&~{\cal N}_q ~|\Phi_{\rm cm}(m_2)|^2~ 
			{\alpha_s\over 2\pi}C_f \ln
			{\Lambda^2\over \mu^2}~ (1-x)(1+x)\nonumber\\
&&~~~~~~~+ ~{\cal N}_q~m_2
   			|\Phi_{\rm cm}(m_2)|^2~\Big\{~\delta(1-x)+
			{\alpha_s\over 2\pi}C_f \ln    	  
                        {\Lambda^2\over \mu^2}~ {x^3+x\over 1-x}~\Big\}\, . 
\end{eqnarray}
Note that eq.(\ref{st1}) and eq.(\ref{st2}) are similar except the first
term having opposite sign and $m_1$ is replaced by $m_2$.   
Adding them together, we obtain the final result as
given in eq.(\ref{qll}).



\begin{references}
\bibitem{rak} A. Harindranath and Rajen Kundu, Phys. Rev. D {\bf 59}, 
116013, (1999).
\bibitem{zhari}  A. Harindranath, Rajen Kundu and W. M. Zhang, Phys. Rev D
{\bf 59}, 094013, (1999). 
\bibitem{schaf} P. H\"{a}gler and A. Sch\"{a}fer, Phys. Lett. B {\bf 430}, 179
(1998).
\bibitem{zhang93} W. M. Zhang and A. Harindranath, Phys. Rev. D {\bf 48},
4881 (1993).
\bibitem{ji} X. Ji, Phys. Rev. Lett. {\bf 78},
610, (1997).
\bibitem{bj} S.V. Bashinsky and R.L. Jaffe, Nucl. Phys. B {\bf 536}, 509,
(1998). 
\bibitem{dz} D. Singleton and V. Dzhunushaliev, hep-ph/9807239.
\end{references}
\end{document}